\documentclass[aps,prl,twocolumn]{revtex4}
\usepackage {amssymb}
\usepackage {amsmath}
\usepackage {graphicx}

\newcommand{\SSC}{\scriptscriptstyle}
\newcommand{\bs}{\boldsymbol}

\begin{document}
\title{Spatially Differential Forms of Lenz Law}
\author{Ji Luo}
\affiliation{Institute of High Energy Physics, Chinese Academy of
Sciences, Beijing, 100039, China}
\date{\today}

\begin{abstract}
Two sets of spatially differential formulas of Lenz law on
electromagnetic inductance are presented. They are a cut-magnetic
flux induced voltage, which instantaneously results from cutting
magnetic flux as a conductor moving with respect to an external
magnetic field, and a wave-induced voltage due to the interaction
of a conductor or charged particles with the arriving
electromagnetic wave, which is originated from a changing magnetic
flux source. Upon the Lenz differential forms, the induced
electrical field strength, relevant properties and their
application are discussed.
\end{abstract}
\maketitle

Lenz law, as an empirical rule, sums production and property of
inductive electromotive force in a conductor loop when an
interaction occurs between the circle loop and a steady magnetic
field or changeable magnetic field. It is expressed in an integral
formula of induced electromotive voltage on whole conductor loop
and can be applied for a totally induced electrical voltage
calculation in steady-symmetric circumstances~\cite{dictofphy},
\begin{equation}
\label{eq1} \varepsilon _t (t) = \oint
 {\bs{E}_t (t) \cdot \mathrm{d} \bs{l}}  =  - \oint_\ell  {\delta \mathit{\dot {\Phi} }_t
 (t)}\;.
\end{equation}

However in physical practice there are variety of induction
phenomena which occur in a non-steady status with asymmetric
circumstances; or the induced electrical strength are needed to be
calculated upon an interestedly particular point, segment of
conductor on differently spatial condition as such where the
integral form \textbf{\textit{Eq.(\ref{eq1})}} \textbf{\textit{can
not be in use}}~\cite{luoji:phdthesis,luoji:energydiff}. Moreover
\textbf{\textit{the integrated form will cover some explicit
details of physical properties that can be revealed and applied
through its spatially differential forms}}. Like the important
role of Biot-Savart theorem (or differential form of Ampere
theorem) on magnetic field
calculation[~\cite{dictofphy,introtophy}, therefore, the
presentation of the differential form of Lenz law will be
significant to time-dependent calculation of induced electrical
field.

Actually electromagnetic inductances occur on a spatial conductor
element or charged particles much more often than on a whole
conductor loop. Mathematically the integral form is merely an
accumulated sum of spatial differential inductance results along
the whole loop. And in viewpoint of physics, the causes of all
electromagnetic inductance phenomena could be divided into two
categories: cut-magnetic flux induced electrical field and
wave-induced electrical field. The integral form is given as
\begin{equation}
\label{eq2}
\begin{split}
 \varepsilon _t (t) & =  \oint
{\bs{E}_{\mathrm{c}} (t) \cdot \mathrm{d}\bs{l}} - \oint {\bs{E}_w (t)\cdot \mathrm{d}\bs{l}} \\
& =  - \left[ {\oint {\delta \mathit{\dot {\Phi } }_{d\ell   c}
(t)} + \oint
{\delta \mathit{\dot {\Phi }}_{d\ell   w} (t)} } \right] \\
& = - \oint {\delta \left[ {\mathit{\dot {\Phi}
 }_{d\ell  c} (t) + \mathit{\dot {\Phi }}_{d\ell   w} (t)} \right]} \;.\\
\end{split}
\end{equation}


Upon Eq.(\ref{eq2}), for having differently physical properties
the cut-magnetic flux induced field and wave-induced field on the
conductor are classified into two categories. The former is
related to energy conversion from mechanical kinetic energy to
electrical energy, and the latter related to the conversion from
an electromagnetic wave kinetic energy to an associated energy
carried or possessed by a conductor or charged particles.

The spatial differential Form of the cut-magnetic field induced
voltage can be given below,
\begin{equation}
\label{eq3}
\begin{split}
 d\varepsilon _c (t)& = - \left[ {\bs{v} _{d\ell \cdot b} (t)\times \bs{B}(t)}
\right] \cdot \mathrm{d}\bs{l} \\
 &= - \bs{E}_c (t) \cdot \mathrm{d}\bs{l} \\
 &= - \delta \mathit{\dot {\Phi } }_{d\ell  c} (t)\;. \\
 \end{split}
\end{equation}

Where: $-\frac{\pi}{2} \leqslant \alpha_c =
\widehat{\bs{n}_{\kern-0.2em\lower0.3ex\hbox{$\SSC{\bs{E}}_c$}}\!
(t) \bs{n}_{d\bs{l}}(t)} \leqslant \tfrac{\pi }{2}$, $0 \leqslant
\delta \mathit{\dot {\Phi }}_{d\ell  c} (t) = \bs{E}_c (t) \cdot
\bs{n}_{dl} (t) d\ell $.

$\bs{v} _{d\ell \cdot b} (t)$ --- relative velocity of conductor
element $d\bs{l}$ with respect to $\bs{B}(t)$, $\bs{v} _{d\ell
\cdot b} = \bs{v} _{d\ell } - \bs{v} _b $;

$\bs{E}_c (t)$ --- induced electrical field strength in conductor
element by cutting magnetic flux; or induced electrical force per
unit length along $\bs{v} _{d\ell \cdot b} (t)\times \bs{B}(t)$
direction by cutting $\bs{B}(t)$ field.

As conductor element $d\bs{l}$ moves in external magnetic field
the Lorentz action will cause an energy conversion from a
non-electrical energy (such as kinetic energy) to an electrical
energy. It is the action that results in an induced electrical
field strength $\bs{E}_c(t)$ inside of the conductor, or an
elemental induced electromotive force or voltage $d\varepsilon _c
(t)$ in conductor, its value depends upon induced electrical field
strength, and size and orientation of the conductor. The
derivation is following:
\begin{widetext}
\begin{equation}
\label{eq4}
\begin{split}
 d\varepsilon _{d\ell   c} \left[ {\bs{r}(t)} \right] &= - B\left[ {\bs{r}(t)}
\right]A_{c \bot b} (t) =-B v_{\bot}d\ell_{_{//\bs{v}\times\bs{B}}} = -Bv_{_{d\ell\cdot b}} \sin\theta_c d\ell \cos\alpha_c  \\
& = \bs{B}(t) \cdot \left\{ {\left[ {\bs{v} _{d\ell \cdot b}
(t)\times \bs{n}_b (t)}
\right]\times \left[ {d\bs{l}\times \bs{n}_b (t)} \right]} \right\} \\
& = - B(t)\left[ {\bs{n}_b (t)\times d\bs{l}} \right] \cdot
\left\{ {\bs{n}_b (t)\times
\left[ {\bs{v} _{d\ell \cdot b} (t)\times \bs{n}_b (t)} \right]} \right\} \\
& = - B(t)\left[ {\bs{n}_b (t)\times d\bs{l}} \right] \cdot
\left\{ {\left[ {\bs{n}_b (t) \cdot \bs{n}_b (t)} \right]\bs{v}
_{d\ell \cdot b} (t) - \left[ {\bs{n}_b (t) \cdot
\bs{v} _{d\ell \cdot b} (t)} \right]\bs{n}_b (t)} \right\} \\
& = - B(t)\left[ {\bs{n}_b (t)\times d\bs{l}} \right] \cdot
\left\{ {\bs{v} _{d\ell \cdot b} (t) - \left[ {\bs{n}_b (t) \cdot
\bs{v} _{d\ell \cdot b} (t)}
\right]\bs{n}_b (t)} \right\} \\
& = - B(t)d\bs{l} \cdot \left\{ {\left\{ {\bs{v} _{d\ell \cdot b}
(t) - \left[ {\bs{n}_b (t) \cdot \bs{v} _{d\ell \cdot b} (t)}
\right]\bs{n}_b (t)} \right\}\times
\bs{n}_b (t)} \right\} \\
& = - B(t)\left[ {\bs{v} _{d\ell \cdot b} (t)\times \bs{n}_b (t)}
\right] \cdot
d\bs{l} \\
& = - \left[ {\bs{v} _{d\ell \cdot b} (t)\times \bs{B}(t)} \right]
\cdot \bs{n}_{d\bs{l}} (t) d\ell \\
& = - \bs{E}_c (t) \cdot \bs{n}_{d\bs{l}} (t) d\ell \\
& = - \delta \mathit{\dot {\Phi }}_{d\ell  c} (t) \;.\\
 \end{split}
\end{equation}
\end{widetext}
Here $\theta_c$: the angle between the direction of the magnetic
flux density at the conductor location and that of $bs{v}_{d\ell
\cdot b}(t)$ or $\theta_c=\widehat{\bs{n}(\bs{v}_{d\ell\cdot
b})\bs{n}_b}$ and
\begin{equation}
\label{eq5} \bs{E}_c (t) = \bs{v} _{d\ell \cdot b} (t)\times
\bs{B}(t) = v_{_{d\ell\cdot b}}B \sin\theta_c \bs{n}(\bs{v}_{d\ell
\cdot b}\times\bs{B})\;.
\end{equation}

Take an integrated form of Eq.(\ref{eq4}) there exists
\begin{equation}
\label{eq6}
\begin{split}
 \varepsilon _c (t) &= \oint_\ell {d\varepsilon _{d\ell  c} (t)} \\
 &= - \oint_\ell {\left[ {\bs{v} _{d\ell \cdot b} (t)\times \bs{B}(t)} \right]
\cdot \mathrm{d}\bs{l}} \\
 &= - \oint_\ell {\bs{E}_c (t) \cdot \mathrm{d}\bs{l}} \\
 &= - \oint_\ell {\delta \mathit{\dot {\Phi }}_{d\ell  c} (t)} \;.\\
 \end{split}
\end{equation}

Eq.(\ref{eq5},\ref{eq6}) reveal that at spatially different points
of the conductor, Lorentz action exerted on the positive and
negative charges within the conductor, the separately
electromotive force in the opposite directions, is the only cause
to form the cut-induced electrical field within a conductor.

Another type of electromagnetic inductance results from the
interaction of the conductor or charged particles as
electromagnetic load with the arriving electromagnetic wave. The
energy of the electromagnetic radiation, originated from a
changing magnetic flux source or a source of magnetic flux rate,
will induce an associated electrical field at the conductor
location and be transferred into an associated energy of the
conductor or charged particles in their overlapped space.
\begin{equation}
\label{eq7}
\begin{split}
 \bs{E}_{sw} ({t}') &= \bs{E}_{sw} ({t}' = t + T) \\
 &= \frac{\bs{n}(\bs{r}_{s d\ell } )\times \bs{n}_b \dot {B}\left[ {\bs{r}_s (t)}
\right]ds_{ \bot b} dL}{4\pi r_{s d\ell }^2 } \\
 &= \frac{\bs{n}(\bs{r}_{s d\ell } )\times \bs{n}_b \delta \mathit{\dot {\Phi}}_w (t)dL}{4\pi
r_{s d\ell }^2 } \\
 &= \frac{\bs{n}(\bs{r}_{s d\ell } )\times \dot {\bs{B}}\left[ {\bs{r}_s (t)}
\right]dV}{4\pi r_{s d\ell }^2 } \\
 &= \frac{\bs{n}(\bs{r}_{s d\ell } )\times \delta \dot {\bs{M}}(t)}{4\pi r_{s d\ell }^2
}\\
&=\frac{\dot{B}\left[\bs{r}_s(t)\right]dV sin \theta(t)}{4\pi
r^2_{sd\ell}}\bs{n}\left[\bs{n}(\bs{r}_{sd\ell})\times\bs{n}_b\right]\\
&=E_w(t')\bs{n}_{_{\bs{E}_w}}(t')\;,
 \end{split}
\end{equation}
where, $\bs{E}_{sw} (t')$ --- the induced electrical field
strength, or induced electrical force per unit length on a
conductor along $\bs{n}(\bs{r}_{s d\ell } )\times \delta \dot
{\bs{M}}(t)$ direction by electromagnetic wave originated from a
micro magnetic flux rate element $\delta \dot {\bs{M}}(t)$ in an
air and vacuum
space;\\
 $\delta \dot {\bs{M}}= \dot {\bs{B}}\left[ {\bs{r}_s (t)}
\right]dV = \bs{n}_b\delta \mathit{\dot {\Phi }}_w (t)dL =
\bs{n}_b \dot {B}\left[ {\bs{r}_s (t)} \right]ds_{ \bot b} dL$ ---
a spatial element of magnetic flux's rate;\\
 $\bs{n}_b $
--- the unit vector of magnetic flux density $\bs{B}\left[ {\bs{r}_s (t)}
\right]$ at $t$ instant and $\bs{r}_s (t)$position;\\
$\theta_w$ --- the angle between the direction of
$\bs{B}[\bs{r}_s(t)]$ and that of the
relative position vector $\bs{r}_{s d\ell}$;\\
$\bs{n}\left[\bs{n}(\bs{r}_{sd\ell})\times\bs{n}_b\right]=\bs{n}\left[\bs{n}(\bs{r}_{sd\ell})\times
\delta \dot{ \bs{M}}(t)\right]=\bs{n}_{_{\bs{E}_w}}(t')$ in an air
and vacuum space;\\
 $0 \leqslant \alpha_w =
\widehat{\bs{n}_{_{\bs{E}_w} } ({t}') \bs{n}_{d\bs{l}} ({t}')}
\leqslant {\pi}$, $ 0 \leqslant \delta \mathit{\dot {\Phi
}}_{d\ell  w} ({t}') = \bs{E}_w ({t}') \cdot \bs{n}_{d\bs{l}}
({t}')d\ell$;\\
 $\bs{r}_{s d\ell } = r_{s d\ell } \bs{n}(\bs{r}_{s
d\ell } )$
--- relative position vector between a micro magnetic flux's rate
spatial element $\delta \dot {\bs{M}}=\dot {\bs{B}}\left[
{\bs{r}_s (t)} \right]dV$ and a micro conductor element
$d\bs{l}$;\\
$t$, ${t}'$ --- emitting instant and arriving instant respectively
of a micro electromagnetic wave element originated from the
magnetic flux's rate source, here ${t}' - t = T(t) = T({t}')$.

So the generally spatial differential form of the wave-induced
voltage will be given below,
\begin{equation}
\label{eq8}
\begin{split}
 d\varepsilon _w ({t}') &= - \left[ {\sum {\bs{E}_{jsw} ({t}')} } \right] \cdot d\bs{l}
\\
& = - \left[ {\sum {\frac{\bs{n}(\bs{r}_{js d\ell } )\times
\bs{n}_b \delta \mathit{\dot {\Phi}
}_{jw} (t_j )dL_j }{4\pi r_{js d\ell }^2 }} } \right] \cdot d\bs{l} \\
& = - \left[ {\sum {\frac{\bs{n}(\bs{r}_{js d\ell } )\times \dot
{\bs{B}}\left[ {\bs{r}_{js} (t_j )} \right]dV_j }{4\pi r_{js d\ell
}^2 }} } \right]
\cdot d\bs{l} \\
& = - \left[ {\sum {\frac{\bs{n}(\bs{r}_{js d\ell } )\times \delta
\dot {\bs{M}}_j
(t_j )}{4\pi r_{js d\ell }^2 }} } \right] \cdot d\bs{l} \\
& = - \bs{E}_w ({t}') \cdot d\bs{l} \\
& = - \delta \mathit{\dot {\Phi }}_{d\ell w} \left[ {\bs{r}_{d\ell
} ({t}')} \right]\;.
 \end{split}
\end{equation}

Eq.(\ref{eq7},\ref{eq8}) manifest a general explicitly formula of
purely wave-induced electrical field strength at a detector
position where the charged particles or a conductor serve as a
electromagnetic load to interact with an arriving electromagnetic
wave. In the process an energy exchange will occur among the
electromagnetic energy possessed by the load and wave
respectively. It is known that the associated induced electrical
strength is related to spatial propagating distribution,
attenuation and time delay of the wave-energy.

For a steady wave-induction circumstance, the induced voltage
element can be expressed:
\begin{equation}
\label{eq9}
\begin{split}
 d\varepsilon _w ({t}') &=  - \bs{E}_w ({t}') \cdot d\bs{l} \\
 &= - \left[ \iiint_{\dot {M}} {\frac{\bs{n}(\bs{r}_{s d\ell}) \times \delta \dot{\bs{M}}(t)}{4\pi r_{s d\ell }^2 }} \right] \cdot d\bs{l} \\
 &= - \left[ \iiint_V {\frac{\bs{n}(\bs{r}_{s d\ell} )\times \dot{\bs{B}}(t)\mathrm{d}V}{4\pi r_{s d\ell }^2 }} \right] \cdot
 d\bs{l}\;.
 \end{split}
\end{equation}

Or when the spatial shape of the magnetic flux's rate source is
long tube like shape with small cross section area $ds_{ \bot b} $
usually taken in practical cases, then the induced electrical
force can be expressed as:
\begin{equation}
\label{eq10}
\begin{split}
 \bs{E}_w ({t}') &= \iiint_{\dot {M}} {\frac{\bs{n}(\bs{r}_{s d\ell } )\times \delta
\dot {\bs{M}}(t)}{4\pi r_{s d\ell }^2 }} \\
 &= \iiint_V {\frac{\bs{n}(\bs{r}_{s d\ell } )\times \dot {\bs{B}}(t)\mathrm{d}V}{4\pi
r_{s d\ell }^2 }} \\
 &= \int_L {\iint_{ds} {\frac{\bs{n}(\bs{r}_{s d\ell } )\times \bs{n}_b \dot
{B}(t)\mathrm{d}s_{ \bot b} }{4\pi r_{s d\ell }^2 }}} \mathrm{d}L \\
 &= \int_L {\frac{\bs{n}(\bs{r}_{s d\ell } )\times \bs{n}_b \delta \mathit{\dot
 {\Phi}
}(t)}{4\pi r_{s d\ell }^2 }} \mathrm{d}L\;.
 \end{split}
\end{equation}

\begin{figure}[tbp]
\centerline{\includegraphics[width=1.26in,height=1.53in]{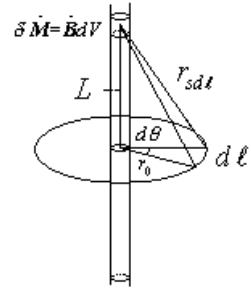}}
\caption{Diagram of wave-induced electrical field by a micro
magnetic flux rate element}\label{fig1}
\end{figure}

Following is an illustrating example of calculating wave-induced
electrical field strength in steady-symmetry circumstance (Ref.
Fig. \ref{fig1}).

Take an integral form of Eq.(\ref{eq10}), there then exists
\begin{equation}
\label{eq11}
\begin{split}
 \bs{E}_w ( {t}' ) &= \iiint_{\dot {M}} {\frac{\bs{n}(\bs{r}_{s d\ell }
)\times \delta \dot {\bs{M}}( {t}' )}{4\pi r_{s d\ell }^2 }} \\
 &= \iiint_V {\frac{\bs{n}(\bs{r}_{s d\ell } )\times \dot {\bs{B}}t( {t}'
)\mathrm{d}V}{4\pi r_{s d\ell }^2 }} \\
 &= \int_L {\frac{\bs{n}(\bs{r}_{s d\ell } )\times \dot {B}\left( {t}'
\right)ds_{ \bot b} \bs{n}_b \mathrm{d}L}{4\pi r_{s d\ell }^2 }} \\
 &= \int_0^\infty {\frac{2r_0 \delta \mathit{\dot {\Phi }}( {t}')\bs{n}_{\kern-0.2em\lower0.3ex\hbox{$\SSC{\bs{E}}_w$}}
}{4\pi (r_0 ^2 + L^2)^{\raise0.5ex\hbox{$\scriptstyle
3$}\kern-0.1em/\kern-0.15em\lower0.25ex\hbox{$\scriptstyle 2$}}}\mathrm{d}L} \\
 &= \frac{\delta \mathit{\dot {\Phi }}( {t}')}{2\pi r_0 }\bs{n}_{\kern-0.2em\lower0.3ex\hbox{$\SSC{\bs{E}}_w$}} = E_w
( {t}')\bs{n}_{\kern-0.2em\lower0.3ex\hbox{$\SSC{\bs{E}}_w$}} \;,\\
 \end{split}
 \end{equation}
it is $\left| {\bs{E}_w ({t}')} \right| = \tfrac{\delta
\mathit{\dot {\Phi }}({t}')}{2\pi r_0 }$. So
\begin{equation*}
\begin{split}
 \varepsilon _w ({t}') &= \oint_\ell {E_w ({t}')\bs{n}_{\kern-0.2em\lower0.3ex\hbox{$\SSC{\bs{E}}_w$}} \cdot \mathrm{d}\bs{l}} =
\oint_\ell {\frac{\delta \mathit{\dot {\Phi }}({t}')}{2\pi r_0 }\bs{n}_{\kern-0.2em\lower0.3ex\hbox{$\SSC{\bs{E}}_w$}} \cdot \mathrm{d}\bs{l}} \\
 &= - \oint_\ell {\frac{\delta \mathit{\dot {\Phi }}({t}')}{2\pi r_0 }\mathrm{d}\ell = -
\int_0^{2\pi } {\frac{\delta \mathit{\dot {\Phi }}({t}')}{2\pi }\mathrm{d}\theta } } \\
 &= - \delta \mathit{\dot {\Phi }}({t}') \;.
 \end{split}
\end{equation*}

Meanwhile by applying Eq.(\ref{eq1}) in the case, the following
equations can be derived.
\begin{equation}
\label{eq12} \varepsilon _w ( {t}') = \oint {d\varepsilon _w (
{t}') = } \oint {E_w ( {t}' ) \cdot d\ell } = 2\pi r_0 E_w ({t}')
= - \delta \mathit{\dot {\Phi }}( {t}')\;,
\end{equation}
and
\begin{equation}
\label{eq13} \left| {\bs{E}_w ({t}')} \right| = \frac{\delta
\mathit{\dot {\Phi }}({t}')}{2\pi r_0 }.
\end{equation}

Eq.(\ref{eq11}, \ref{eq12}) testifies that the calculated results
are consistent by using any one among the differential form of
wave-induced electrical field and Lenz integrated form along a
closed circular loop in a steady-symmetry induction situation with
even-radial electromagnetic radiant wave from center of the loop.



Combining the two types of inductance together, there are the
consequences on the total inductance forms as following.
\begin{equation}
\label{eq14}
\begin{split}
 &\bs{E}_t ({t}') = \bs{E}_c ({t}') + \bs{E}_w ({t}') \\
&= \bs{v} _{d\ell \cdot b} ({t}')\times \bs{B}({t}') + \sum
{\frac{\bs{n}(\bs{r}_{js d\ell } )\times \delta \dot {\bs{M}}_j
({t}' - T_j )}{4\pi
r_{js d\ell }^2 }} \\
&= \bs{v} _{d\ell \cdot b} ({t}')\times \bs{B}({t}') + \sum
{\frac{\bs{n}(\bs{r}_{js d\ell } )\times \dot {\bs{B}}_j \left[
{\bs{r}_s \left( {{t}' - T_j } \right)} \right]dV_j }{4\pi r_{js
d\ell }^2 }}\;.
 \end{split}
\end{equation}

For steady electromagnetic induction situation, an induced
electrical strength at a point can be expressed as below,
\begin{equation}
\label{eq15}
\begin{split}
 \bs{E}_t ({t}') &= \bs{E}_c ({t}') + \bs{E}_w ({t}') \\
 &= \bs{v} _{d\ell \cdot b} ({t}')\times \bs{B}({t}') + \iiint_{\dot{M}}
{\frac{\bs{n}(\bs{r}_{s d\ell } )\times \delta \dot
{\bs{M}}(\bs{r}_s ,{t}')}{4\pi
r_{s d\ell }^2 }} \\
 &= \bs{v} _{d\ell \cdot b} ({t}')\times \bs{B}({t}') + \iiint_V
{\frac{\bs{n}(\bs{r}_{s d\ell } )\times \dot {\bs{B}}(\bs{r}_s
,{t}') dV}{4\pi r_{s d\ell }^2 }}\;.
 \end{split}
\end{equation}

The total induced voltage on the conductor element is given then,
\begin{equation}
\label{eq16}
\begin{split}
 d\varepsilon _t ({t}') &= - \bs{E}_t ({t}') \cdot d\bs{l} \\
 &= - \left[ {\bs{E}_c ({t}') + \bs{E}_w ({t}')} \right] \cdot d\bs{l} \\
 &= - \delta \left[ {\mathit{\dot {\Phi }}_{d\ell  c} ({t}') + \mathit{\dot {\Phi }}_{d\ell  w}
({t}')} \right]\;.
 \end{split}
\end{equation}

The spatial differential forms can be applied mainly to two
aspects.

\begin{enumerate}
\item Calculation of induced electrical field strength will be
conducted on \textbf{\textit{any particular point, segment or
non-closed loop of conductor with asymmetry circumstance }}where
Lenz integral form can not be applied in purpose, such as the
calculation of wave-induced electrical field strength at center
point of a toroid source of changing magnetic flux; the
calculation of wave-induced electrical field ($\bs{E}_w )$ could
be applied to a time-dependent spatial ($\bs{E}_w )$ field
distribution.

\item Explicit and quantitative analysis on the energy conversion,
spatial orientation or polarization of induced electrical field
strength on a conductor element and their time rate in a dynamic
process of electromagnetic inductance. In addition, the time delay
of the wave-induced signal, due to the wave propagating from the
source of magnetic flux rate to the conductor element, can be
analyzed upon the spatial differential form.
\end{enumerate}


Lenz experimental law can be reduced to two fundamentally spatial
differential forms: cut-magnetic field induced electrical field
and wave-induced electrical field on a conductor element. In both
cases a relative velocity between the conductor and a contact
magnetic field or magnetic flux rate field is necessary
preposition for inductance occurrence. \textbf{\textit{The spatial
differential forms not only manifest explicitly the physical
essence of electromagnetic inductance but also can be applied as
the calculating formulas in variety of calculation and analysis of
electromagnetic inductance phenomena}} where Lenz integral form
can \textbf{\textit{not}} offer the relevant explicit solution as
its mathematical limit.

\begin{acknowledgments}
The author would like to thank Bo Liu and Wenchun Gao for
intensive discussions.
\end{acknowledgments}


\begin{thebibliography}{4}
\expandafter\ifx\csname
natexlab\endcsname\relax\def\natexlab#1{#1}\fi
\expandafter\ifx\csname bibnamefont\endcsname\relax
  \def\bibnamefont#1{#1}\fi
\expandafter\ifx\csname bibfnamefont\endcsname\relax
  \def\bibfnamefont#1{#1}\fi
\expandafter\ifx\csname citenamefont\endcsname\relax
  \def\citenamefont#1{#1}\fi
\expandafter\ifx\csname url\endcsname\relax
  \def\url#1{\texttt{#1}}\fi
\expandafter\ifx\csname
urlprefix\endcsname\relax\def\urlprefix{URL }\fi
\providecommand{\bibinfo}[2]{#2}
\providecommand{\eprint}[2][]{\url{#2}}

\bibitem[{\citenamefont{Gray and Isaacs}(1975)}]{dictofphy}
\bibinfo{author}{\bibfnamefont{H.~J.} \bibnamefont{Gray}} \bibnamefont{and}
  \bibinfo{author}{\bibfnamefont{A.}~\bibnamefont{Isaacs}},
  \emph{\bibinfo{title}{A New Dictionary of Physics}}
  (\bibinfo{publisher}{Longman}, \bibinfo{address}{New York},
  \bibinfo{year}{1975}), pp. \bibinfo{pages}{19--20,173--174}.

\bibitem[{\citenamefont{Luo}(2000)}]{luoji:phdthesis}
\bibinfo{author}{\bibfnamefont{J.}~\bibnamefont{Luo}},
  \emph{\bibinfo{title}{Electromagnetic response of transmission line under
  short circuit by carbon-based fiber}}, \bibinfo{howpublished}{Ph.D.
  Dissertation, Beijing Institute of Technology, Beijing}
  (\bibinfo{year}{2000}).

\bibitem[{\citenamefont{Luo}(2003)}]{luoji:energydiff}
\bibinfo{author}{\bibfnamefont{J.}~\bibnamefont{Luo}},
  \emph{\bibinfo{title}{Energy differential structure and energy exchange of a
  micro element of charged particles in longintudinal acceleration}},
  \bibinfo{howpublished}{Postdoctoral Research Report, Institute of High Energy
  Physics, Chinese Academy of Sciences, Beijing} (\bibinfo{year}{2003}).

\bibitem[{\citenamefont{Bueche}(1975)}]{introtophy}
\bibinfo{author}{\bibfnamefont{F.~J.} \bibnamefont{Bueche}},
  \emph{\bibinfo{title}{Introduction to physics for scientists and engineers}}
  (\bibinfo{publisher}{McGraw-Hill Book Company}, \bibinfo{address}{New York},
  \bibinfo{year}{1975}), pp. \bibinfo{pages}{456--472, 475--497}.

\end{thebibliography}

\end{document}